\documentclass[aps,pra,superscriptaddress,twocolumn]{revtex4}
\usepackage{bm}
\usepackage{graphicx}
\usepackage{amsmath}
\usepackage{amssymb}
\usepackage{txfonts}

\begin{document}

\newcommand {\mean}[1]{\langle #1 \rangle}
\newcommand {\meantwo}[1]{\langle #1 \rangle}
\newcommand{\parall}{\uparrow\uparrow}
\newcommand{\antiparall}{\uparrow\downarrow}
\newcommand\lavg{\left\langle}
\newcommand\ravg{\right\rangle}
\newcommand\ket[1]{\left|#1\right\rangle}
\newcommand\bra[1]{\left\langle#1\right|}

\title{Manipulating and protecting entanglement by means of spin environments}

\author{T. J. G.  Apollaro}
\affiliation{Dipartimento di Fisica, Universit\`a di Firenze,
    Via G. Sansone 1, I-50019 Sesto Fiorentino (FI), Italy}

\author{A. Cuccoli}
\affiliation{Dipartimento di Fisica, Universit\`a di Firenze,
    Via G. Sansone 1, I-50019 Sesto Fiorentino (FI), Italy}
\affiliation{CNISM - Consorzio Nazionale Interuniversitario per le Scienze Fisiche della Materia}

\author{C. Di Franco}
\affiliation{Department of Physics, University College Cork, Republic of Ireland}

\author{M. Paternostro}
\affiliation{School of Mathematics and Physics, Queen's
University, Belfast BT7 1NN, United Kingdom}

\author{F. Plastina}
\affiliation{{Dipartimento di  Fisica, Universit\`a della Calabria, 87036 Arcavacata di Rende
(CS) Italy}}
\affiliation{{INFN - Gruppo collegato di Cosenza, 87036 Arcavacata di Rende
(CS) Italy}}

\author{P. Verrucchi}
\affiliation{Dipartimento di Fisica, Universit\`a di Firenze,
    Via G. Sansone 1, I-50019 Sesto Fiorentino (FI), Italy}
\affiliation{{ISC - Consiglio Nazionale delle Ricerche, U.O. Dipartimento di Fisica, via
G. Sansone 1, I-50019 Sesto Fiorentino (FI), Italy}}
\affiliation{{INFN Sezione di Firenze,
via G.Sansone 1, I-50019 Sesto Fiorentino (FI), Italy}}
\date{\today}

\begin{abstract}
We study the dynamical behavior of two initially entangled qubits, each
locally coupled to an environment embodied by an interacting spin chain. We consider energy-exchange qubit-environment couplings resulting in a rich and highly non trivial
entanglement dynamics. We obtain exact results for the time-evolution of
the concurrence between the two qubits and find that, by tuning the
interaction parameters, one can freeze the dynamics of entanglement, therefore inhibiting its relaxation into the spin environments, as well as activate a sudden-death phenomenon. We also discuss the effects of an environmental quantum phase transition on the features of the two-qubit entanglement dynamics.
\end{abstract}

\pacs{ 75.10.Pq,05.70.Jk,03.67.-a}

\maketitle

\section{Introduction}

The interplay between coherent and incoherent processes is key
in the quantum mechanical processing of information. Systems designed
in order to perform a given communication or computational task
have to cope with the detrimental effects of the surrounding world, which
can affect an otherwise coherent process in many distinct
ways~\cite{nielsen}.
In the context of distributed quantum information processing (QIP),
where networks of spatially remote quantum nodes are
used in order to process information in a delocalized way, a good
assumption is that each local processor is affected by its own
environment. Such an architecture for a QIP device is currently at the
focus of extensive and multifaceted theoretical and experimental
endeavors~\cite{browne}.

Exstensive work has been performed on the
incoherent dynamics resulting from the coupling of QIP systems with
baths weakly perturbed by the system-induced back-action.
The loss of quantum correlations due to the environment
has recently received considerable
attention~\cite{konrad}, with particular emphasis on the phenomenon
of environment-induced entanglement sudden death~\cite{yu} ({\it
i.e.}, complete disentanglement in a finite time), which has also
been experimentally tested for the case of electromagnetic
environments~\cite{experiments}. And yet, especially for
solid-state implementations of quantum processors, the case of
structured environments is extremely relevant~\cite{spinenv}. In this framework, the dynamics
leading to complete disentanglement of two qubits coupled with a
common spin environments (the so-called ``central-qubit model") has been extensively
studied~\cite{zi,sun,cecilia}. Exponential decay of
the concurrence~\cite{wootters} between two qubits initially prepared in pure
states has been observed, the decay rate
being enhanced for working points close to
the quantum phase transition of the environment~\cite{sun}.

In this paper, we study the dynamical evolution of the entanglement
between two remote qubits coupled with mutually independent spin
environments. The exact time dependence of their reduced density
matrix is obtained using an original
approach~\cite{DiFrancoEtal07,DiFrancoEtal08} which allows us to
track the dynamics of quantum averages of one- and two-spin
observables and entanglement properties. In stark contrast with most
of the available literature, here we consider a ``transversal''
qubit-environment coupling that allows for energy transfer, resulting
in a dissipative-like behavior of the qubits and in a much richer
entanglement dynamics.

The transversal nature of the qubit-environment coupling allows us to
reveal the occurrence of entanglement sudden death even when starting
from initially pure states of the qubits, there included the
maximally entangled ones which are extremely relevant for QIP. The
above feature cannot emerge in the case of longitudinal couplings as
reported in Refs.~\cite{zi,sun,cecilia,davide}.

We show that, by setting the environment in different operating
regime, one can induce either entanglement sudden death or a freezing
effect. Moreover, we shed light onto the differences in behavior
experienced by the {\it parallel} and {\it antiparallel} concurrence
introduced in~\cite{FubiniEtal06}, whose interplay is crucial in
determining the properties of the quantum correlations within the
two-qubit state, in particular at the environmental quantum critical
point where the antiparallel entanglement appears to be better
preserved than the parallel one.

The paper is organized as follows. In Sec.~\ref{s.system} we define
the physical system and the relevant interactions. In
Sec.~\ref{s.dynamicsI} we provide exact expressions for the dynamics
of a single qubit coupled with a spin-environment. These results are
used in Sec.~\ref{s.dynamicsII} in order to get the dynamics of the
entanglement between two qubits, each coupled with a spin-environment
via a local isotropic (Subsection~\ref{ss.isocoupling}) or
anisotropic (Subsection~\ref{ss.anisocoupling}) exchange interaction.
Finally, in Sec.~\ref{s.conclusions} we draw our conclusions and
suggest a possible scenario where the main features of our physical
model can be embodied.

\section{The system}
\label{s.system}

\begin{figure}
\includegraphics[width=0.55\linewidth]{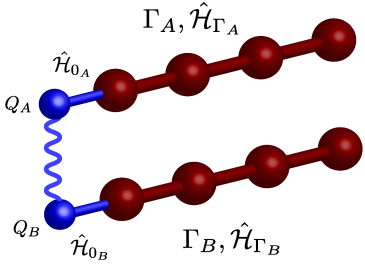}
\caption{Sketch of the physical model. Each of a pair of entangled qubits, $Q_A$ and $Q_B$, is locally coupled
with a spin chain, $\Gamma_A$ and $\Gamma_B$, via the Hamiltonians $\hat{\cal H}_{0_A}$ and $\hat{\cal H}_{0_B}$, respectively. The dynamics within each chain is ruled by the intra-chain Hamiltonians,
$\hat{\cal H}_{\Gamma_A}$ and $\hat{\cal H}_{\Gamma_B}$.
The wavy line indicates initial entanglement.}
\label{f.system}
\end{figure}

We consider two non-interacting subsystems, $A$ and $B$, each
consisting of a qubit $Q_\kappa$ ($\kappa=A,B$) coupled with a chain
$\Gamma_\kappa$ of $N_\kappa$ interacting $S=1/2$ particles.
Whenever useful we will hereafter use the index $\kappa=A,B$
so as to generically refer to either of the two subsystems $A$ or $B$.
As usual, the qubits $Q_\kappa$ are described in terms of
$S=1/2$ spin operators, which we indicate as
${\hat{\bm s}}_{0\kappa}$. Operator ${\hat{\bm s}}_{n_\kappa}$ ($n_\kappa=1,...,N_\kappa$)
corresponds to the spin located at site $n$ of the chain $\Gamma_\kappa$.
Notice that although the above notation suggests the spin describing
$Q_\kappa$ to sit at site $0$ of the respective chain $\Gamma_\kappa$,
this is just a useful convention, but has no implication about the physical
nature of $Q_A$ and $Q_B$.

The intra-chain interaction is of $XY$-Heisenberg
type with local magnetic fields possibly applied in the $z$-direction
\begin{equation}
\label{e.HGamma_k}
\hat{\cal{H}}_{\Gamma_{\kappa}}\!=\!-\!2\!\sum_{n_\kappa=1}^{N_\kappa-1}
(J^x_{n_\kappa} \hat{s}^x_{n_\kappa} \hat{s}^x_{n_\kappa+1}\!+\!
 J^y_{n_\kappa} \hat{s}^y_{n_\kappa} \hat{s}^y_{n_\kappa+1})
-2\sum_{n_\kappa=1}^{N_\kappa} h_{n_\kappa} \hat{s}^z_{n_\kappa},
\end{equation}
where $h_{n_\kappa}$ is the field applied at site $n_\kappa$ and
$J^{x,y}_{n_\kappa}$ are the coupling strengths of the intra-chain
interactions.
Each $\Gamma_\kappa$ is open-ended, while neither the
$J^{x,y}_{n_\kappa}$'s nor the magnetic fields $h_{n_\kappa}$ need to be uniform
along the chains. Qubit $Q_\kappa$ is coupled with the first spin of its environment, embodied by $\Gamma_\kappa$, via an exchange
interaction of strengths $J_{0_\kappa}^{x,y}$ and can be subjected
to a local magnetic field $h_{0_\kappa}$ directed along the $z$-direction. The corresponding Hamiltonian reads
\begin{equation}
\label{e.H_0k}
\hat{\cal{H}}_{0_\kappa}\!=\!-2(
J^x_{0_\kappa}\hat{s}^x_{0_\kappa}\hat{s}^x_{1_\kappa}+
J^y_{0_\kappa}\hat{s}^y_{0_\kappa}\hat{s}^y_{1_\kappa})\!-\!
2h_{0_\kappa}\hat{s}^z_{0_\kappa}~.
\end{equation}
The Hamiltonian of the total
system $\kappa$ is thus given by
$\hat{\cal H}_{\kappa}=\hat{\cal H}_{0_\kappa}+\hat{\cal
H}_{\Gamma_\kappa}$. In
Fig.~\ref{f.system} we provide a sketch of the model considered.
As stated previously, although $A$ and $B$ do not interact
directly, they experience joint dynamics due to the possibility
of sharing initial entanglement. Depending on the choice of the local interaction parameters and magnetic fields in
$\hat{\cal{H}}_\kappa$ the efforts required to tackle the model
greatly changes.
In Section~\ref{s.dynamicsI} we describe the approach used in order to
achieve an exact solution for the dynamics of local observables.

\section{Exact single-qubit dynamics}
\label{s.dynamicsI}

We first concentrate on the dynamics of one subsystem only (thus
dropping the index $\kappa$ throughout this Section). In
particular, we are interested in the evolution of a given initial state of
the qubit $Q$, as determined by its coupling with the chain $\Gamma$ and under
the influence of the local magnetic field. We resort to the Heisenberg
picture, which has been recently shown to
provide a convenient framework for the analysis of quantum many-body systems
of interacting particles~\cite{DiFrancoEtal07}. The key step is
the use of the Campbell-Baker-Hausdorff  (CBH) formula in the management of the time-evolution operator of the system. For an operator $\hat{\cal{O}}$ associated with an observable of a physical system with Hamiltonian $\hat{\cal{H}}$, the CBH formula reads (we set $\hbar=1$ throughout the paper)
\begin{equation}
\label{e.BCH}
\hat{\cal{O}}(t)= \sum_{p=0}^\infty\frac{(i t)^p}{p!}
\left[\hat{\cal{H}},\left[\hat{\cal{H}},..
\left[\hat{\cal{H}},{\hat{\cal O}}\right]..\right]\right].
\end{equation}
 In virtue of
the algebra satisfied by the Pauli matrices, we find that upon application of Eq.~(\ref{e.BCH}), the time evolution of the components
of ${\hat{\bm s}}_0$ reads
\begin{equation}
\label{e.XYZ0(t)}
\begin{aligned}
\hat{s}^x_0(t)&=\frac{1}{2}\sum_{n=0}^N
\left[\Pi^x_n(t)\hat{\sigma}^x_n+\Delta^x_n(t)\hat{\sigma}^y_n\right]
\hat\sigma_0^z\hat P_n~,
\\
\hat{s}^y_0(t)&=\frac{1}{2}\sum_{n=0}^N
\left[\Pi^y_n(t)\hat{\sigma}^y_n-\Delta^y_n(t)\hat{\sigma}^x_n\right]
\hat\sigma_0^z\hat P_n~,
\\
\hat{s}^z_0(t)&=-i2\hat{s}^x_0(t)\hat{s}^y_0(t),
\end{aligned}
\end{equation}
where $\hat{\sigma}_n^\alpha$ ($\alpha=x,y,z$) are the Pauli
operators for the spin at site $n$ and
$\hat P_n=\prod_{i=1}^{n-1}\hat{\sigma}^z_i$.
The time-dependent coefficients $\Pi_n^{x}(t)$
and $\Delta_n^{x}(t)$ are the components of the $(N+1)$-dimensional
vectors ${\bm\Pi}^{x}(t)$ and ${\bm\Delta}^{x}(t)$ defined by
\begin{eqnarray}
{\bm\Pi}^x(t)&=&\sum_{p=0}^{\infty}(-1)^p\frac{t^{2p}}{(2p)!}
({\bm \tau}{\bm \tau}^T)^p{\bm v},
\label{e.PiX}\\
{\bm\Delta}^x(t)&=&\sum_{p=0}^{\infty}(-1)^{p}\frac{t^{2p+1}}{(2p+1)!}
{\bm \tau^T}({\bm \tau}{\bm \tau}^{T})^p{\bm v},
\label{e.DeltaX}
\end{eqnarray}
where $T$ stands for transposition, the vector ${\bm v}$ has components
$v_i=\delta_{i0}$~\cite{commentoordering} and
the tri-diagonal matrix ${\bm \tau}$ has elements
\begin{equation}
\tau_{ij}=
J_{i-1}^y\delta_{i-1,j}+J_i^x\delta_{i+1,j}-2 h_i\delta_{i,j}~.
\label{e.T}
\end{equation}
The coefficients ${\bm \Pi}^{y}(t)$
and ${\bm \Delta}^{y}(t)$ are obtained from
Eqs.~(\ref{e.PiX}) and (\ref{e.DeltaX}) by replacing
${\bm \tau}$ with ${\bm \tau}^T$. As $\bm{\tau\tau}^T$ is real and symmetric, there is an orthogonal
matrix ${\bm U}$ that diagonalizes it, so that
$(\bm{\tau\tau}^T)^p={\bm U}{\bm\Lambda}^{2p}{\bm U}^T$,
with ${\bm \Lambda}$ the diagonal matrix whose elements
$\lambda_{ij}=\lambda_i\delta_{i,j}$ are the (positive) square roots of
the eigenvalues of ${\bm \tau\bm \tau}^T$.
Similarly, there is an orthogonal matrix ${\bm V}$ that
diagonalizes $\bm{\tau}^T\bm\tau$ and such that
$(\bm{\tau}^T{\bm\tau})^p={\bm V}{\bm\Lambda}^{2p}{\bm V}^T$
with the same diagonal matrix ${\bm\Lambda}$ as above.
As a consequence, after straightforward matrix algebra, one can
sum up the time-dependent
series in Eqs.~(\ref{e.PiX}) and (\ref{e.DeltaX}) to get
\begin{equation}
\label{e.PiDeltaxyResum}
\begin{aligned}
{\bm \Pi}^x(t)&={\bm U}{\bm \Omega}(t){\bm U}^T{\bm v},~~
{\bm \Delta}^x(t)={\bm V}{\bm\Sigma}(t){\bm U}^T{\bm v}\\
{\bm \Pi}^y(t)&={\bm V}{\bm \Omega}(t){\bm V}^T{\bm v},~~
{\bm \Delta}^y(t)={\bm U}{\bm\Sigma}(t){\bm V}^T{\bm v},
\end{aligned}
\end{equation}
where ${\bm \Omega}(t)$ and ${\bm \Sigma}(t)$ are diagonal
matrices with elements
$\Omega_{ij}(t)=\cos(\lambda_i t) \delta_{ij}$ and
$\Sigma_{ij}(t)=\sin(\lambda_i t) \delta_{ij}$.
The above equations hold regardless of the
local magnetic fields or the couplings $J^{x,y}_n$ and $J^{x,y}_{0}$.
Adopting the language of
Refs.~\cite{DiFrancoEtal07,DiFrancoEtal08}, the components of ${\bm
\Pi}^{x,y}(t)$ and ${\bm \Delta}^{x,y}(t)$ embody the {\it fluxes of
information} from the qubit $Q$ to the spin chain
$\Gamma$.

By means of Eqs.~(\ref{e.XYZ0(t)}) one can determine the time
evolution of the single-qubit density matrix
\begin{equation}
\label{e.rho0}
\hat{\rho}_0(t)= \frac{1}{2}\hat{\openone}+\sum_{\alpha}
\mean{\hat{s}^\alpha_0(t)}\hat{\sigma}^\alpha_0,
\end{equation}
where $\hat{\openone}$ is the identity operator and
$\mean{\,\cdot\,}$ indicates the expectation value over the initial
state of the system. Once the diagonalizations required for
determining the vectors ${\bm\Pi}^{x(y)}(t)$ and
${\bm\Delta}^{x(y)}(t)$ are performed, we need to evaluate the
expectation values entering Eq.~(\ref{e.rho0}). Such a task can be
performed within two different scenarios. In the first, $\Gamma$
has a small number of spins, so that one can {\it
design} the precise structure of its initial state. In the second,
$\Gamma$ consists of a large number of spins, which puts the analysis in the thermodynamic limit, where we can benefit of specific
symmetry properties of $\hat{\cal{H}}_\Gamma$.
Here, we concentrate on the latter
situation, which has been the subject of several recent papers,
due to the fact that it can be used for describing a proper
qubit-environment system.

We assume that $Q$ and $\Gamma$ are initially uncorrelated and set the former in an arbitrary single-qubit
state $\hat{\rho}_0$ and the latter in an eigenstate
$\ket{\Psi_\Gamma}$ of $\hat{\cal{H}}_\Gamma$. The initial state of
the total system will
thus be
$\hat{\rho}=\hat{\rho}_0\otimes\ket{\Psi_\Gamma}\bra{\Psi_\Gamma}$.
Under such conditions, the procedure described in the above section is most
conveniently implemented as Eqs.~(\ref{e.XYZ0(t)})
greatly simplify due to the properties of
$\hat{\cal{H}}_\Gamma$. In particular, the conservation rule
${[\hat{\cal{H}},\bigotimes_{n=0}^{N}\hat\sigma^z_{n}]=0}$, which states
parity invariance
of the Hamiltonian, together with the property
$\mean{\hat\sigma_n^\alpha\hat\sigma_m^\beta}=0$
for $\alpha\neq \beta$ and $n\neq m$, imply
\begin{equation}
\mean{\hat s^x_0(t)}{=}\frac{1}{2}\sum_{n=0}^N
\mean{\hat\sigma_0^z}
\left[\Pi^x_n(t)\mean{\hat P_n\hat\sigma^x_n}
+\Delta^x_n(t)\mean{\hat P_n\sigma^y_n}\right]
\label{e.meanX0(t)}
\end{equation}
\begin{equation}
\mean{\hat s^y_0(t)}{=}\frac{1}{2}\sum_{n=0}^N
\mean{\hat\sigma_0^z}
\left[\Pi^y_n(t)\mean{\hat P_n\hat\sigma^y_n}-
\Delta^y_n(t)\mean{\hat P_n\hat\sigma^x_n}\right]
\label{e.meanY0(t)}
\end{equation}
\begin{equation}
\begin{aligned}
&\mean{\hat s^z_0(t)}{=}
\frac{1}{2}\sum_{n=0}^N
\left[\Pi_n^x(t)\Pi_n^y(t) + \Delta_n^x(t)\Delta_n^y(t)\right]
\mean{\hat\sigma_n^z}\label{e.meanZ0(t)}\\
-&\frac{1}{2}\sum_{n<m}^N
\left[\Pi_n^y(t) \Pi_m^x(t) + \Delta_n^x(t) \Delta_m^y(t)\right]
\mean{\hat P_{n+1}\hat P_m\hat\sigma_n^x\hat\sigma_m^x}\\
-&\frac{1}{2}\sum_{n<m}^N
\left[\Pi_n^x(t) \Pi_m^y(t) + \Delta_n^y(t) \Delta_m^x(t)\right]
\mean{\hat P_{n+1}\hat P_m\hat\sigma_n^y\hat\sigma_m^y}~.
\end{aligned}
\end{equation}
Using Eqs.~(\ref{e.rho0})-(\ref{e.meanZ0(t)}) one can finally evaluate the
single-qubit density matrix $\hat{\rho}_0(t)$.

\section{Exact two-qubit dynamics}
\label{s.dynamicsII}
We now consider the complete system $A{\cup}B$. Under the
assumption of non-interacting subsystems, the time propagator
generated by $\hat{\cal H}_A+\hat{\cal H}_B$ factorizes as
$\hat{\cal{U}}_A(t)\otimes\hat{\cal{U}}_B(t)$
with $\hat{\cal{U}}_\kappa(t)=\exp[-i\hat{\cal{H}}_\kappa t]$.
Despite the absence of interaction, $A$ and $B$ might still display dynamical
correlations depending on the initial state of the total system. In
fact, if $A{\cup}B$ is prepared in an entangled state,
its dynamical properties will depend on the structure of such
initial state and the entanglement evolution will follow from the interactions ruling $A$ and $B$
separately.
The values of the parameters entering
$\hat{\cal{H}}_{\kappa}$ might thus
be considered as {\it knobs} for the entanglement dynamics.

We prepare the total system at time $t=0$ in
\begin{equation}
\hat\rho(0)=
\hat\rho^{Q_A Q_B}(0)\otimes
\hat\rho^{\Gamma_A}(0)\otimes\hat\rho^{\Gamma_B}(0),
\label{e.initialState}
\end{equation}
so that we can use the results of Sec.~\ref{s.dynamicsI} and write the
two-qubit density matrix at time $t$ in terms of the time-evolved
single-qubit one. In fact, using the results of Ref.~\cite{BellomoEtal07}, we have
\begin{equation}
{\rho}^{Q_\kappa}_{ij}(t)=
\sum_{p,r}K^{pr}_{ij}(t){\rho}^{Q_\kappa}_{pr}(0)~~~~~(\kappa,K=A,B)
\label{e.rho1Q}
\end{equation}
with $\rho^{Q_\kappa}_{ij}(t)$ the elements of the single-qubit density matrix
at time $t$ and ${\bm K}$ a tensor of time-dependent coefficients.
The two-qubit state can then be written as
\begin{equation}
\label{e.rho2Q}
{\rho}^{Q_A Q_B}_{IJ}(t)=
\sum_{p_A,r_A}\sum_{p_B,r_B}A_{i_A j_A}^{p_A r_A}(t)B_{i_B j_B}^{p_B
r_B}(t){\rho}^{Q_A Q_B}_{PR}(0),
\end{equation}
where lower-case indices take values $1$ and $2$, while capital
ones are defined according to $L=l_A+l_B-l_A{\rm mod}2=1,..,4$
(with $L=I,J,P,R$ and $l=i,j,p,r$).
By setting
$\hat\rho_{\Gamma_\kappa}(0)=
\ket{\Psi_{\Gamma_\kappa}}\bra{\Psi_{\Gamma_\kappa}}$,
{\it i.e.} by preparing both the chains in any of the eigenstate of their
respective Hamiltonians, we can explicitely evaluate
Eqs.~(\ref{e.meanX0(t)})-(\ref{e.meanZ0(t)}) and hence Eq.~(\ref{e.rho0}).
By comparing the latter with Eq.~(\ref{e.rho1Q}) we finally determine the
coefficients $K^{pr}_{ij}(t)$, thus fully specifying the dynamics of the two-qubit
state $\hat\rho^{Q_A Q_B}(t)$.

Our approach is fully general and can be used in a variety of different situations.
Here we concentrate on the case where the initial state of the two qubits is
one of the Bell states
\begin{equation}
\label{e.initialStateXX}
\ket{\phi_\pm}=\frac{1}{\sqrt 2}\left(\ket{00} \pm \ket{11}\right),~~~~
\ket{\psi_\pm}=\frac{1}{\sqrt 2}\left(\ket{01} \pm \ket{10}\right),
\end{equation}
which we dub {\it parallel}
($\ket{\phi_\pm}$) and {\it antiparallel} ($\ket{\psi_\pm}$) Bell
states~\cite{FubiniEtal06}.
In virtue of the symmetries of such states,
the concurrence of the two-qubit state can be written as
${C(t){=}2\max\{0,C_{\parall}(t),C_{\antiparall}(t)\}}$, where we have introduced
\begin{eqnarray}
C_{\antiparall}(t)&{=}&
|\tilde\rho_{23}|{-}\sqrt{\tilde\rho_{11}\tilde\rho_{44}}~,\label{e.Cantiparallel}\\
C_{\parall}(t)&{=}&|\tilde\rho_{14}|{-}\sqrt{\tilde\rho_{22}\tilde\rho_{33}}~\label{e.Cparallel}
\end{eqnarray}
with $C_{\parall}(0)=1$ for
$\hat\rho_0^{Q_AQ_B}(0)=\ket{\phi_\pm}\bra{\phi_\pm}$
and $C_{\antiparall}(0)=1$ for
$\hat\rho_0^{Q_AQ_B}(0)=\ket{\psi_\pm}\bra{\psi_\pm}$.
Here the notation $\tilde\rho_{IJ}$ is used, for the sake of clarity, to
indicate the matrix elements $\rho_{IJ}^{Q_A Q_B}(t)$ defined in
Eq.~(\ref{e.rho2Q}). In what follows $C_{a}(t)$ [$C_p(t)$] is the concurrence corresponding to
the case where the qubits are initially prepared in an antiparallel [parallel] Bell state.

As for the environments, we take two identical chains of an (equal) even number of spins N
with homogeneous intra-chain couplings and field, {\it i.e.}  $J^{x,y}_{n_\kappa}{=}J$  and $h_{n_\kappa}{=}h$. Both chains are prepared
in the ground state of ${\cal{H}}_{\Gamma_\kappa}$, which is found via
Jordan-Wigner and Fourier transformations~\cite{sachdev,pla}.
Straightforward calculations yield the relevant mean values entering
Eqs.~(\ref{e.XYZ0(t)}) as
\begin{equation}
\begin{aligned}
\mean{\hat\sigma_n^z}&{=}1{-}\frac{2}{N+1}
\left(k_F-\frac{\cos[(k_F+1)\vartheta_n]\sin[k_F \vartheta_n]}{\sin\vartheta_n}\right),\\
g_{nm}&{\equiv}\mean{\hat P_{n}\hat P_m\hat\sigma_n^x\hat\sigma_m^x}=
\frac{\phi_{n,k_F+1}\varphi_{m,k_F}-\varphi_{n,k_F}\varphi_{m,k_F+1}}
{2\left(\cos\vartheta_n-\cos\vartheta_m\right)},
\end{aligned}
\end{equation}
where, for convenience of notation, we omit the index $\kappa$. We have introduced
$\varphi_{j,k}{=}\sqrt{{2}/{(N+1)}}\sin(j \vartheta_k)$,
${\vartheta_k={k\pi}/({N+1})}$, ${k\in [1,N]}$ and the Fermi
wave number $k_F$ is determined by the magnetic field~\cite{pla}.
Moreover, due to the absence of symmetry breaking in
$\ket{\Psi_{\Gamma _\kappa}}$, it is
$\mean{\hat\sigma^{x,y}_{n_\kappa}}=0$. We consider $N$ finite but large enough to avoid finite-size effects
to influence our results. For the range of  parameters
considered here, $N=50$  is found to fulfill such condition and it is then chosen to set the length of
both chains. Finally, we take the same coupling strength between each qubit and its
chain, that is $J^\alpha_{0_\kappa}=J^\alpha_0$.

\subsection{Isotropic coupling between the qubit and the chain}
\label{ss.isocoupling}

We now consider an isotropic coupling between each qubit
and its respective environment, defined by
$J^x_{0_\kappa}=J^y_{0_\kappa}=J_0$ in Eq.~(\ref{e.H_0k})
In the theory of open quantum systems, such coupling typically corresponds
to a dissipative interaction treated in the rotating wave approximation
\cite{heinzpeter}.
For isotropic coupling, the total Hamiltonians $\hat{\cal{H}}_\kappa$ have
rotational
invariance along the $z$-axis. This implies $\bm\tau=\bm\tau^{T}$,
and thus $\Pi_n^x(t)=\Pi_n^y(t)\equiv\Pi_n(t)$,
$\Delta_n^x(t)=\Delta_n^y(t)\equiv\Delta_n(t)$, which allows us to write
\begin{equation}
\label{meanopsum}
\begin{aligned}
&\mean{\hat s^x_0(t)}= \frac{1}{2}(\Pi_0(t) \mean{\hat\sigma^x_0} +
\Delta_0(t) \mean{\hat\sigma^y_0}),\\
&\mean{\hat s^y_0(t)}=\frac{1}{2}(\Pi_0(t)\mean{\hat\sigma^y_0} -
\Delta_0(t) \mean{\hat\sigma^x_0}),\\
&\mean{\hat s^z_0(t)}=\frac{1}{2}
\sum_{n=0}^{N} [\Pi^2_n(t) + \Delta^2_n(t)]\mean{\hat\sigma^z_n}\\
&-\frac{1}{4}\!\sum_{\substack{n\neq{m}=1}}^N
[\Pi_n(t)\Pi_m(t)+\Delta_n(t)\Delta_m(t)]g_{nm}~.\\
\end{aligned}
\end{equation}
From the above expressions we see that single-qubit states
initially directed along the $z$-axis of the Bloch-sphere maintain such
alignment regardless of the Hamiltonian parameters. On the other hand,
initial ``equatorial'' states with $\mean{\hat s^z_0(0)}=0$ evolve in
time and remain on the equatorial plane only for zero overall magnetic
field.
The rotational invariance around the $z$-axis of the total Hamiltonians
$\hat{\cal{H}}_A$ and $\hat{\cal{H}}_B$ has relevant consequences also
on the evolution of the entanglement, as shown in Section~\ref{ss.anisocoupling}.

In the fully homogeneous case, {\it i.e.} for $h_0=h$
and $J_0=J$, it is $\lambda_q=-2(h-\cos\frac{q\pi}{N+2})$
and the $j^{th}$ component of the corresponding eigenvector has the same
form as $\varphi_{j,q}$ with $N+1$ replaced
by $N+2$. Hereafter, time will be measured in units of $J^{-1}$.
In the thermodynamic limit where $N\rightarrow\infty$,
the summations can be replaced by integrals yielding
\begin{equation}
\label{exactxy0}
\begin{aligned}
\mean{\hat s^x_0(t)}&=\frac{{\cal J}_1(2t)}{2
t}\left[\cos(2ht)\mean{\hat\sigma^x_0}-\sin(2ht)\mean{\hat\sigma^y_0}\right],\\
\mean{\hat s^y_0(t)}&=\frac{{\cal J}_1(2t)}{2
t}\left[\cos(2ht)\mean{\hat\sigma^y_0}+\sin(2ht)\mean{\hat\sigma^x_0}\right],
\end{aligned}
\end{equation}
and, for the $z$ component
\begin{equation}
\label{exactz0}
\begin{aligned}
&\mean{\hat s^z_0(t)}=\sum_{n=0}^N \frac{(n+1)^2}{2t^2}{\cal J}^2_{n+1}(2t)
\mean{\hat\sigma^z_n}\\
&-\sum_{n\neq m}^{n,m \, even}(-1)^{\frac{n+m}{2}} \frac{(n+1)(m+1)}{4t^2}
{\cal J}_{n+1}(2 t){\cal J}_{m+1}(2 t) g_{nm}\\
&+\sum_{n\neq m}^{n,m \,
odd}(-1)^{\frac{n+m}{2}} \frac{(n+1)(m+1)}{4t^2}
{\cal J}_{n+1}(2 t){\cal J}_{m+1}(2 t) g_{nm},
\end{aligned}
\end{equation}
where ${\cal J}_n(x)$ are the Bessel functions. Long-time expansions show that
the mean value of the single-spin $x (y)$ component decays as $t^{-\frac{3}{2}}$.
If ${h_0=h=0}$, Eq.~(\ref{exactz0}) reduces to $\mean{s^z_0(t)}=\gamma(t)
\mean{\hat\sigma^z_0}$ with $\gamma(t)={{\cal J}^2_{1}(2t)}/{2 t^2}$, yielding a $t^{-3}$
scaling at long times.
Eqs.~(\ref{exactxy0}) and (\ref{exactz0}) give an excellent approximation also for
finite $N$ ($\lesssim50$) within a time range where finite-size effects have not
yet occurred. As the latter are caused by reflection of propagating
excitations at the boundary of the chain, we can neglect finite-size effects for
times up to $\sim{N}$. In fact, as the
maximum one-excitation group velocity is $2$, it takes at least a
time $N$ for an excitation to leave $Q$ and travel back to
it.

Let us now analyze the evolution of the entanglement between the two
qubits. We first notice that in the present case of isotropic coupling
between $Q_\kappa$ and $\Gamma_\kappa$ and given that the exchange
interaction along the chain is set to be of XX type, both Hamiltonians $\hat{\cal H}_\kappa$
have rotational invariance around the $z$ axis. This
enforces a disjoint dynamics of the off-diagonal matrix elements
($\tilde\rho_{23}$ and $\tilde\rho_{14}$) entering
Eq.~(\ref{e.Cantiparallel}) and (\ref{e.Cparallel}). As a consequence, if
$C_{\antiparall}(0)=0$ it is $C_{\antiparall}(t)=0$ at any time
[the same holds for $C_{\parall}(t)$]. Moreover,  we find that $C_{a}(t){\geq}
C_{p}(t)$ for any symmetric ${\bm \tau}$, irrespective of the Hamiltonian parameters. This means
that antiparallel entanglement is more resilient to the effects of the spin environment. Using the analytical solutions given above we finally
obtain the time evolution of the concurrence.
For $h{=}0$ and in the fully omogeneous case of $J_0{=}J$, we find
\begin{equation}
C(t)=2\max\left\{ 0,\gamma^2(t)+\gamma(t)-\frac{1}{2}\right\}
\end{equation}
from which, by using the definition of $\gamma(t)$ in terms of
Bessel functions, we infer that at $t_{\text{ESD}} \simeq 0.9037$ ESD occurs.
In fact, our results show that sudden death occurs also for
$J_{0}\ne J$, exhibiting coupling-dependent characteristics. It is remarkable that
disentanglement at finite time occurs, here, for pure initial states of the two qubits,
a feature due to the specific form of qubit-environment coupling considered here.

\begin{figure}[t]
\includegraphics[width=.7\linewidth]{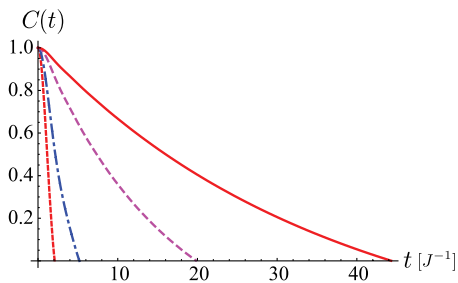}
\caption{Concurrence between $Q_A$ and $Q_B$ for isotropic
and weak coupling, $J_0/J =0.5,0.4,0.2$ and $0.125$ (going from left-most to right-most curve). No magnetic field is applied.}
\label{f.weak}
\end{figure}

\begin{figure}[b]
\includegraphics[width=0.65\linewidth]{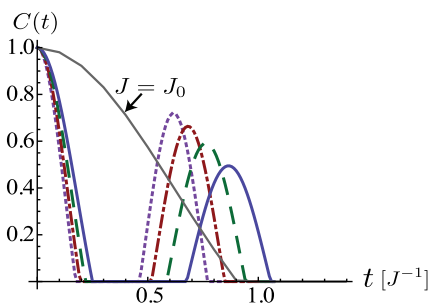}
\caption{Concurrence between $Q_A$ and $Q_B$ for isotropic
and strong coupling, $J_0/J=3.5,4,4.5$ and $5$ (solid, dashed, dot-dashed
and dotted curve, respectively). No magnetic field is applied. We also show the case of $J_0/J=1$}
\label{f.strong}
\end{figure}

\begin{figure}[t]
{\bf (a)}\hskip3.5cm{\bf (b)}
\includegraphics[width=.5\linewidth]{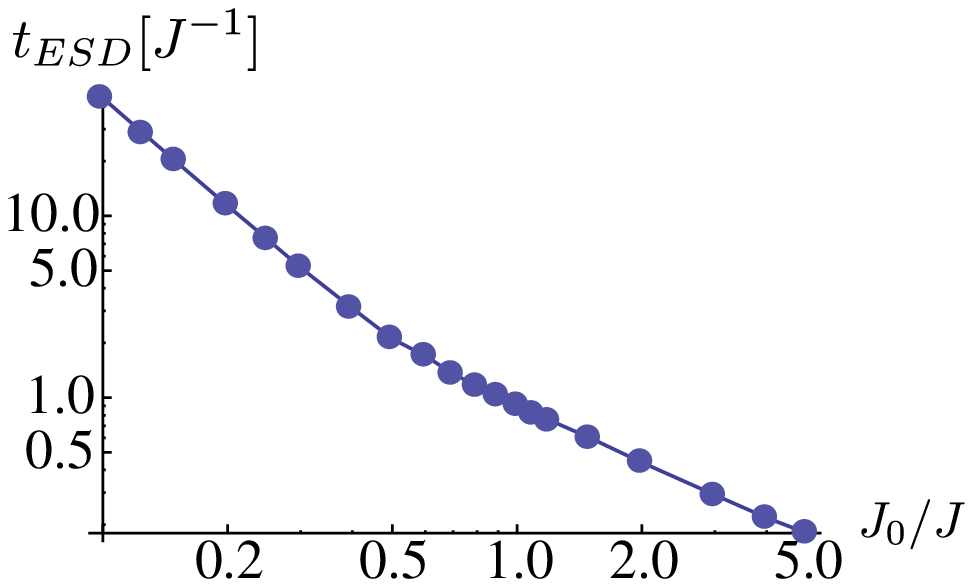}\includegraphics[width=.52\linewidth]{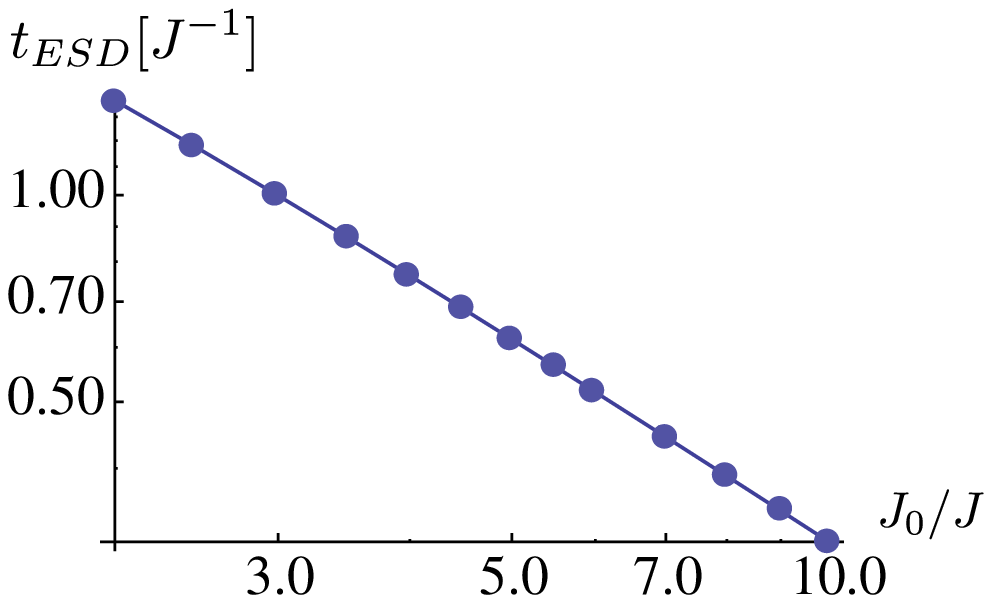}
\caption{{\bf (a)} Time $t_{\rm ESD}$ after which the concurrence between $Q_A$ and
$Q_B$ first exactly vanishes (entanglement sudden-death), as a function of
the coupling $J_0/J$ (double logarithmic scale). No magnetic field is applied. {\bf (b)} Entanglement revival time against $\log_{10}[J_0/J]$. No magnetic field is applied.}
\label{f.tESD}
\end{figure}

In the weak coupling regime $J_0 \ll J$, Fig.~\ref{f.weak} shows that
the concurrence relaxation-time grows as $J_0$ decreases. On the
other hand, for strong coupling $J_0\gg J$, the non-Markovian
character of the environments becomes evident [as shown in
Fig.~\ref{f.strong}]  and entanglement revivals occur due to the
finite memory-time of the spin chain. These revivals can be
intuitively understood as the result of the strong coupling between
the two-qubit system and the spins at the first site of each chain. A
large $J_0$ gives rise to almost perfectly coherent interactions
within such qubit-spin pair, only weakly damped by leakage into the
rest of the chains. A revival time (the time after which the revival
exhibits a maximum) of about $\pi/ J_0$ is inferred from
Fig.~\ref{f.strong}, for $J_0/J \gg 1$. In Fig.~\ref{f.tESD} {\bf
(a)} such findings are summarized in terms of the dependence of
$\log_{10}[t_{\text{ESD}}]$ on $\log_{10}[J_0/J]$. The quasi-linear
trend shown there reveals that the growing rate of $t_{ESD}$ for
$J_0/J<1$ is slightly larger than the decreasing rate at $J_0/J>1$.
In Fig.~\ref{f.tESD} {\bf (b)} we show the behavior of the revival
time $\log_{10}[t_{\text{rev}}]$ against $\log_{10}[J_0/J]$, which
also exhibits a quasi-linear trend.

\begin{figure}[b]
\includegraphics[width=0.495\linewidth]{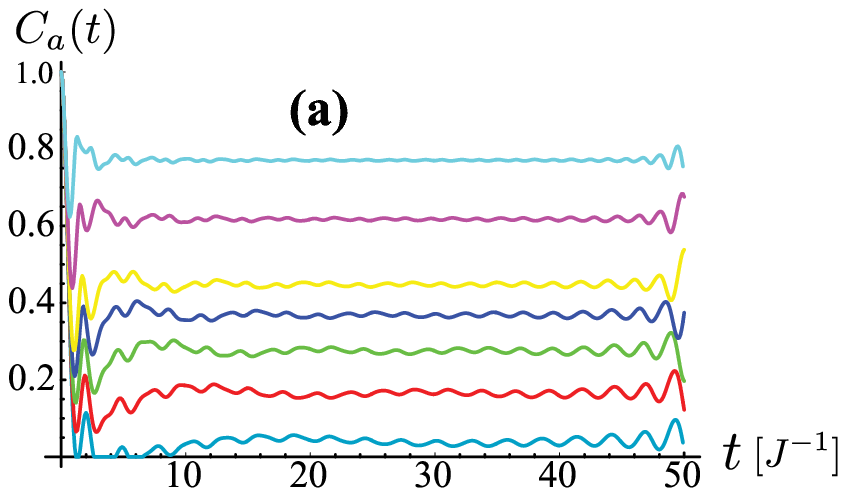}
\includegraphics[width=0.495\linewidth]{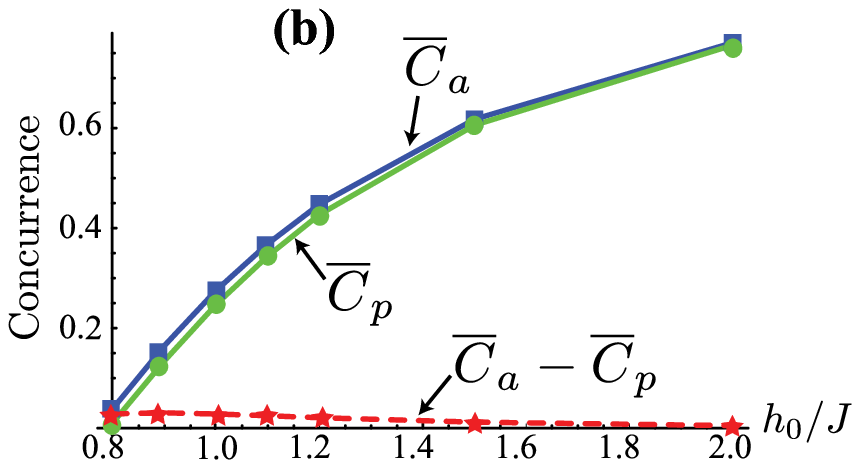}
\caption
{{\bf (a)} $C_{a}(t)$ versus $t$, for isotropic and omogenous
coupling ($J_0{=}J$) on the chains  with $h_{\kappa}{=}0$ and
$h_{0}/J{=}0.8, 0.9, 1, 1.1, 1.2, 1.5, 2$ (from bottom to top curve).
{\bf (b)} Average parallel and
anti-parallel concurrences and their difference plotted
against $h_0$ with the same parameters as in
panel {\bf (a)}. The lines joining the data points are simply a guide to the eye.}
\label{f.iso-omo-fieldonqubits}
\end{figure}

\begin{figure}[hb]
\center{\includegraphics[width=0.495\linewidth]{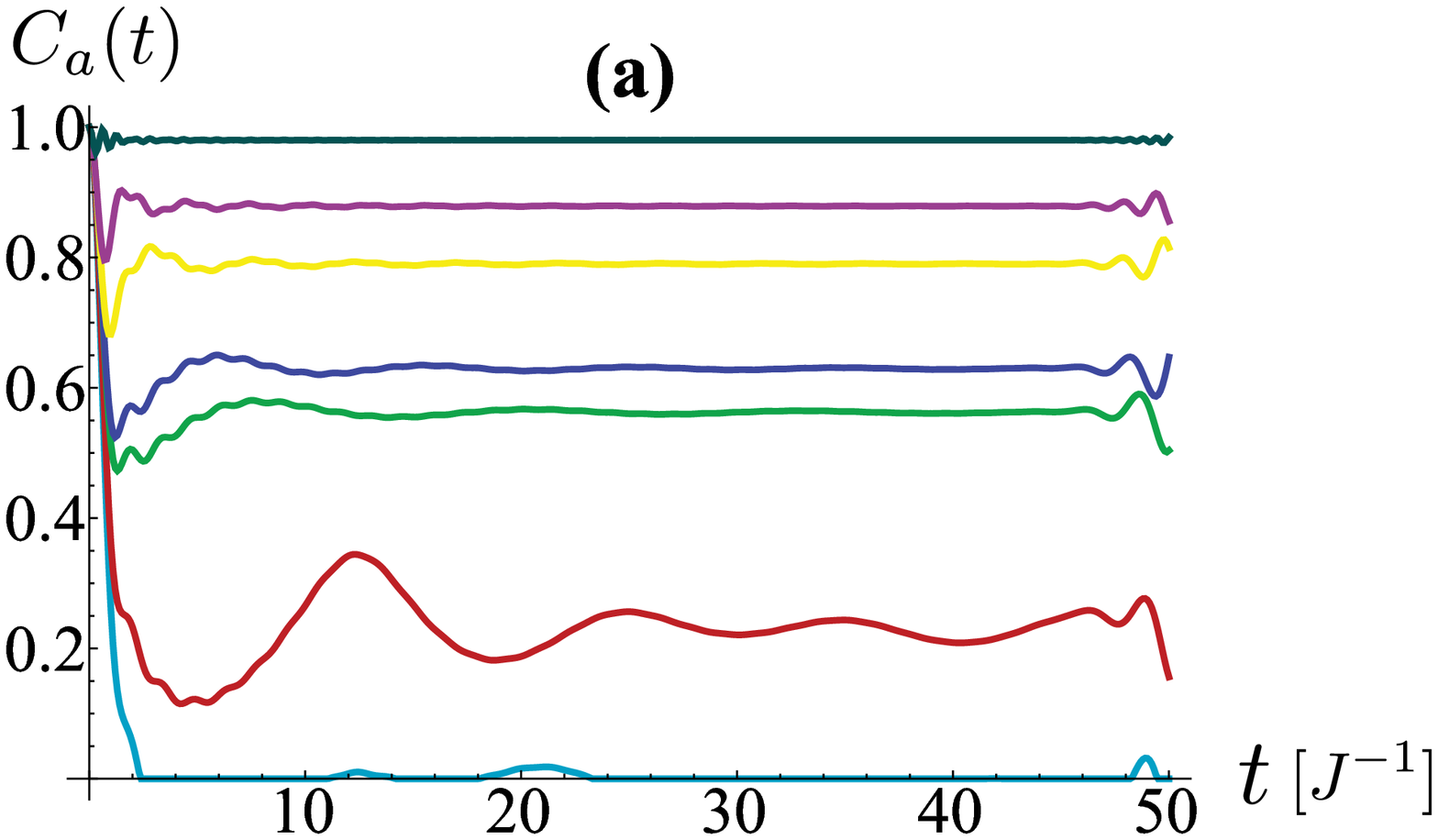}
\includegraphics[width=0.495\linewidth]{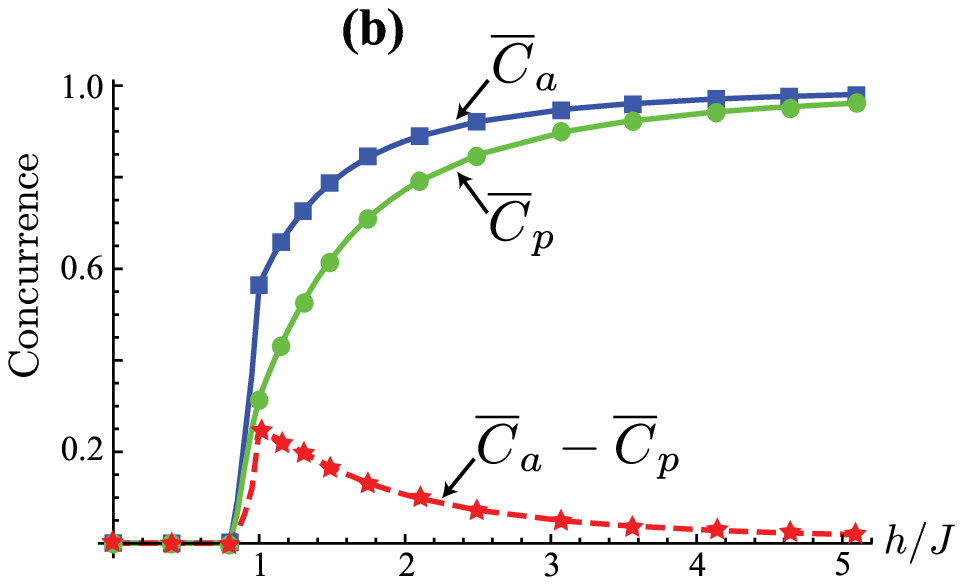}}
\caption
{{\bf (a)}
$C_{a}(t)$ versus $t$, for isotropic and omogenous
coupling ($J_0{=}J$), no field applied on the qubits ($h_{0_{\kappa}}{=}0$) and
$h/J=0.8,0.9,1,1.1,1.5,2,$ and $5$ (from bottom to top curve).
{\bf (b)} Average parallel and
anti-parallel concurrences and their difference plotted
against $h/J$ with the same parameters as in
panel {\bf (a)}. The lines joining the data points are simply a guide to the eye. }
\label{f.iso-omo-fieldonchains}
\end{figure}

The presence of finite magnetic fields significantly changes the dynamics
of the entanglement.
If $h_{0}>0$ and $h=0$, {\it i.e.} the field is only applied to
the qubits, we expect an effective decoupling of
each qubit from the dynamics of its environment, such that
both $\hat{\bm s}_{0_\kappa}$ precess with a Larmor frequency
that depends mostly on $h_{0}$, though it is subjected to
small quantitative corrections due to the
interaction with $\Gamma_\kappa$ at rate $J_{0}$.
In fact, in Fig.~\ref{f.iso-omo-fieldonqubits} we see that
a larger amount of entanglement is mantained for considerably longer times
as $h_0$ grows. Due to the above decoupling, as well as the condition
$h_{0_A}=h_{0_B}$, correlations between $Q_A$ and
$Q_B$ are preserved and so is their concurrence.

In Fig.~\ref{f.iso-omo-fieldonqubits} {\bf (b)} the same effect is
illustrated by the time-averaged concurrence
${\overline{C}_{a,p}=
(1/\delta{t})\int_{\delta{t}}{C}_{a,p}(t')dt'}$ (the average is calculated over a time window $\delta{t}$ that
excludes the oscillatory transients observed in
Fig.~\ref{f.iso-omo-fieldonqubits} {\bf(a)} for $Jt\lesssim{10}$ and
$Jt\gtrsim{45}$). The average entanglement grows with $h_0$.
We further notice that parallel and antiparallel concurrences are almost
identical (with $\overline{C}_a>\overline{C}_p$ as expected),
though their difference vanishes only as $h_0\to\infty$.
We have also found that for $h_{0_A}\neq h_{0_B}$ the phase relation
between individual precessions is lost and no entanglement
preservation is consequently observed.

We now switch off the magnetic field on both qubits (that is, we take $h_0=0$) and
apply a finite field $h>0$ on the environments.
In this case a particularly interesting effect is observed as $h$ becomes
larger than the saturation value $h=J$ and the dynamics of both
chains slows down.
As a consequence, after the transient, the dynamics of the correlations
between the two qubits is considerably suppressed, due to the difficulty
of the qubits to exchange excitations with saturated environments. A long-time
entanglement memory effect results from this, which is evident in
Fig.~\ref{f.iso-omo-fieldonchains} {\bf (a)}. There, we also notice a
reduction of the wiggling, which further witnesses the freezing of the entanglement
dynamics. It should be remarked that such effect is profoundly different from the decoupling mechanism
highlighted previously, where
$\overline{C}_{a}{-}\overline{C}_{p}$ was a
monotonic function of the magnetic field. Here, in fact, a peak
occurs in the difference between time-averaged concurrence
components when $h{=}J$, as shown in Fig.~(\ref{f.iso-omo-fieldonchains}) {\bf (b)},
revealing a drastic change in the entanglement behavior at the onset
of an environmental QPT~\cite{Amico06}. Clearly, at the environmental critical point, the antiparallel entanglement is favored against the parallel one, which is at the origin of the peculiar behavior observed in Fig.~\ref{f.iso-omo-fieldonchains} {\bf (b)} for the dashed line. The drastic change in the behavior of the average concurrence observed at $h/J=1$ is unique of the mechanism discussed here and, as already stressed, clearly distinguished from the freezing effect due to mismatched frequencies at each qubit-environment subsystem. For $h{>}J$, the
effect is fully established and the average concurrence increases, while
$\overline{C}_{a}$ and $\overline{C}_{p}$ get closer to
each other. Moreover, by defining ${\cal{Z}}(t)=(1/2t^2)\sum_{n=1}^N{(n+1)^2 {\cal J}^2_{n+1}(2t)}$ and using the exact analytical expressions (valid for
${h>J}$)
\begin{equation}
\begin{aligned}
C_{a}(t)&{=2\max\left\{0 , \gamma(t){-}
\sqrt{\frac{1}{16}{-}\frac{ \left[
\gamma^2(t){+}{\cal{Z}}^2(t)\right]}{2}+\left[\gamma^2(t){-}{\cal{Z}}^2(t)\right]^2}\right\}},\\
C_{p}(t)&= 2 \max \left\{ 0 , \gamma(t) +
\gamma^2(t) + {\cal{Z}}^2(t)-\frac{1}{4}\right\},
\end{aligned}
\end{equation}
we see that
when the environments are saturated ({\it i.e.} all the spins of the chains are
aligned along the $z$ axis) the concurrence dynamics  does
not depend on the magnetic field.

\subsection{Anisotropic coupling}
\label{ss.anisocoupling}

We finally consider the case of anisotropic coupling $J^x_{0}\neq
J^y_0$ between each qubit and its respective environment (the
chain). Differently from the case of isotropic coupling studied in
Subsection~\ref{ss.isocoupling} and as a direct consequence of the fact that the total
magnetization of $A$ and $B$ is not conserved, the
off-diagonal elements $\tilde{\rho}_{23}$ and $\tilde{\rho}_{14}$ of the
two-qubit reduced density matrix are not dynamically disjoint. This
implies the possibility for the concurrence of the two-qubit state to
switch from the parallel to the antiparallel type and viceversa.

\begin{figure}[t]
\includegraphics[width=0.7\linewidth]{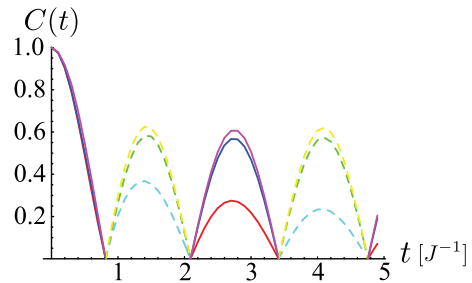}
\caption{Concurrence between $Q_A$ and $Q_B$ for anisotropic coupling,
$J_{0_A}^x=J_{0_B}^x=J$. The magnetic fields are set to zero everywhere
but on the chain $\Gamma_A$, where the field is let to change
within the saturation region, $h_A=1,2,10$ (bottom to top). Solid (Dashed) lines are for for $C_{\parall}$ ($C_{\antiparall}$).
All quantities are dimensionless.}
\label{f.aniso}
\end{figure}

For the sake of clarity, we consider extremely anisotropic conditions, setting ${J_{0_\kappa}^y=0}$.
In the case of no magnetic field on both qubits, ${h_{0}=0}$,
a very simple expression for the concurrence is found, due to the fact
that $\mean{\hat s_{0_\kappa}^x(t)}$ is a constant of motion.
In particular, if the two qubits are initially prepared in a combination of the two antiparallel Bell state, their concurrence will
evolve as ${C_{\antiparall}(t)=-C_{\parall}(t)=\Pi^{y}_{0_A}(t)\Pi^{y}_{0_B}(t)}$.
On the other hand, if parallel Bell states are used to build up the
initial entangled state, it is
${C_{\parall}(t)=-C_{\antiparall}(t)=\Pi^{y}_{0_A}(t)\Pi^{y}_{0_B}(t)}$.
As a consequence, if $\Pi^{y}_{0_A}(t)=\Pi^{y}_{0_B}(t)$ (holding when $\hat{\cal{H}}_A=\hat{\cal{H}}_B$), the two-qubit concurrence cannot
switch between $C_{\parall}$ and $C_{\antiparall}$.
On the contrary, if $\Pi^{y}_{0_A}(t)\neq{\Pi^{y}_{0_B}(t)}$, one can
drive the two-qubit system from parallel-type to
antiparallel and viceversa.
In fact, the switching between parallel and antiparallel
entanglement is observed when $\hat{\bm s}_{0_A}$ and $\hat{\bm s}_{0_B}$
are flipped, under the effect of the coupling with the first spin of their
respective chain, at different frequencies (for instance when $J^x_A\neq
J^x_B$), or when the dynamics of one subsystem is slowed down with respect
to the other (for instance due to the fact that the field on one of the two
chains is larger than the saturation value, as seen in
Sec.~\ref{ss.isocoupling}).
However, as discussed in Ref.~\cite{konrad}, a two-channel
entanglement evolution has an upper bound given by the product of
the one-channel dynamics. Therefore, for such an ``entanglement
switching'' to occur, the Hamiltonian parameters of subsystems $A$
and $B$ should be set so as to retain high entanglement values.
By fixing $J_B^x$, while setting $h_A\gg J^x_A$ in order to slow down the entanglement relaxation in the
corresponding channel, the efficiency of the switching increases.
This is clearly seen in Fig.~\ref{f.aniso}. On the other hand, we
can decrease $J^x_A$ so that channel $B$ is far more responsible
for the entanglement evolution. In this case too a very efficient switching mechanism is achieved, suggesting that the saturation region of the chain is associated to an effective decoupling of the qubit from its corresponding environment. Finally, we notice that, being the coefficients defined by
Eqs.~(\ref{e.PiDeltaxyResum}) regular oscillating functions of time, the concurrence
can only vanish on  a countable number of points on the temporal axis and
cannot remain null for  finite intervals of time. Therefore, entanglement
sudden death is not observed.

\section{Conclusions}
\label{s.conclusions}

We have analyzed the dynamics of an entangled qubit-pair connected to two structured environments
composed of open-ended and finite interacting spin chains.
The intra-chain interaction has been modeled by an XX Heisenberg-like
Hamiltonian, while the coupling between each qubit and its respective environment
has been realized via an XY exchange term with the first spin
of the chain. Application of uniform magnetic fields has been also
considered.
We have exactly determined the time-dependent two-qubit
density matrix, starting from information gathered on the single-qubit
dynamics.

We have then provided analytical solutions both for the case of
finite even $N$ and in the thermodynamic limit, thus getting access
to a full-comprehensive and general analysis of entanglement
evolution. Particular emphasis has been given to the relaxation-like
dynamics implied by the specific type of coupling considered, which
gives rise, under suitable conditions, to a sudden death of the
entanglement that we have analyzed. Interestingly, we have unveiled
the occurrence of ESD also when starting from initially pure
two-qubit states, a peculiarity of our model that does not emerge
under ``longitudinal" qubit-environment couplings.

By manipulating the transverse magnetic field on the initially
entangled qubits, we have shown the possibility of decoupling their
dynamics from that of their respective environments, thereby allowing for
a dynamical entanglement protection.
On the other hand, when the magnetic fields applied to both chains are
larger than the saturation value, the dynamics of the environments slows
down and entanglement sudden death is not observed.
Interesting features are observed when the environments undergo a quantum
phase transition, which in this case happens when the field applied to the
chain gets the saturation value, in particular as far as the the behaviour
of parallel and antiparallel concurrence is concerned.

Our work provides the analytical characterization of the
transverse ({\it i.e.} energy exchanging) coupling between a
simple and yet interesting out-of-equilibrium system (the two qubits) and
a non-trivial spin environments (the two chains).
As spin models are now understood
as effective descriptions of many different physical systems, our
results hold the premises to find fertile application to a variety
of cases. As a particularly interesting situation, one can think
about the engineering of an effective spin environment by using
unidimensional arrays of small-capacitance superconducting
Josephson junctions~\cite{haviland}, which show a sharp phase
transition from Josephson-type behavior to Cooper-pair tunneling
Coulomb blockade analogous to that of an $XY$ model. This
implementation thus constitutes a nearly ideal test for our
predictions, since the effective environmental parameters can be
modified through the use of gate voltages and external magnetic
fluxes. It would be very interesting to study the applicability and relevance
of a study such as the one performed here to the investigation of the
properties of intrinsically open systems in quantum chemistry and
biology exposed to finite-sized environments. In this context, it is
particularly significant that the mathematical model used in order to
describe the radical pair mechanism~\cite{CaiEtAl10} bears some
analogies with the central-qubit model.

\section{Acknowledgements}
We acknoweldge discussions with thank L. Banchi and G. De Chiara. TJGA
thanks Fondazione Carical for financial support. CDF is supported by the Irish Research Council for Science, Engineering and Technology. MP is supported by EPSRC
(EP/G004579/1). MP and FP acknowledge support by the British Council/MIUR British-Italian Partnership Programme 2009-2010.


\begin{thebibliography}{00}
\bibitem{nielsen} M. A. Nielsen and I. L. Chuang, {\it Quantum Computation and Quantum Information} (Cambridge University Press, 2000).

\bibitem{browne} Special issue of Int. J. Quant. Inf. on {\it Distributed quantum computing}, D. E. Browne and S. C. Benjamin (eds.), to appear.

\bibitem{konrad} T. Konrad, F. de Melo, M. Tiersch, C. Kasztelan, A. Arag$\tilde{\text a}$o, and A. Buchleitner,
Nature Phys. {\bf 4}, 99 (2008).

\bibitem{yu} T. Yu and J. H. Eberly, Science {\bf 323}, 598
(2009).

\bibitem{experiments}
M. P. Almeida, F. de Melo, M. Hor-Meyll, A. Salles, S. P. Walborn, P. H. Souto Riberiro, and L. Davidovich, Science {\bf 316}, 579 (2007); J.
Laurat, K. S. Choi, H. Deng, C. W. Chou, and H. J. Kimble, Phys. Rev. Lett. {\bf 99}, 180504 (2007).

\bibitem{spinenv}
C.-Y. Lai, J.-T. Hung, C.-Y. Mou, and P. Chen, Phys. Rev. B {\bf 77}, 205419 (2008); Nie
Jing, Wang Lin-Cheng, and Yi Xue-Xi, Comm. Theor. Phys., {\bf 51},
815 (2009); W. Yang and R.-B. Liu, Phys. Rev. B {\bf 78},
085315 (2008); S. Yuan, M. I. Katsnelson, and H. De Raedt, Phys.
Rev. B {\bf 77}, 184301 (2008).
\bibitem{zi} Z.-G. Yuan, P. Zhang, and S.-S. Li, Phys. Rev. A {\bf 76}, 042118 (2007)

\bibitem{sun} Z. Sun, X. Wang, and C. P. Sun, Phys. Rev. A {\bf 75}, 062312 (2007).

\bibitem{cecilia} C. Cormick and J. P. Paz,
Phys. Rev. A {\bf 78}, 012357  (2008).

\bibitem{wootters}  S. Hill and W. K. Wootters, Phys. Rev. Lett. {\bf 78}, 5022
(1997); W. K. Wootters, Phys. Rev. Lett. {\bf 80}, 2245 (1998).

\bibitem{davide}
D. Rossini, T. Calarco, V. Giovannetti, S. Montangero, and R. Fazio, Phys. Rev. A {\bf 75}, 032333 (2007); D.
Rossini, P. Facchi, R. Fazio, G. Florio, D. A. Lidar, S. Pascazio, F. Plastina, and P. Zanardi, Phys. Rev. A {\bf 77}, 052112 (2008).
\bibitem{DiFrancoEtal07} C. Di Franco, M. Paternostro, G. M. Palma, and M. S. Kim,
Phys. Rev. A {\bf 76}, 042316  (2007).

\bibitem{DiFrancoEtal08} C. Di Franco, M. Paternostro, and M. S. Kim,
Phys. Rev. A {\bf 77}, 020303(R)  (2008).

\bibitem{commentoordering} In order to establish a clear relation with the site-ordering
in the physical model at hand, we label columns and rows of $(N+1)\times (N+1)$ matrices and $(N+1)$-dimensional vectors involved in our formalism using indices ranging
from $0$ to $N$.

\bibitem{BellomoEtal07}
B. Bellomo, R. Lo Franco, and G. Compagno,  Phys. Rev. Lett. {\bf
99}, 160502  (2007).

\bibitem{DiFrancoEtAl08IJQI} C. Di Franco, M. Paternostro, and G. M. Palma,
Int. J. Quant. Inf. {\bf 6}, Supp. 1, 659 (2008).

\bibitem{FubiniEtal06}
A. Fubini, T. Roscilde, V. Tognetti, M. Tusa, and P. Verrucchi,
Eur. Phys. J. D {\bf 38}, 563 (2006).


\bibitem{sachdev} S. Sachdev, {\it Quantum Phase Transitions} (Cambridge University Press, 1999).

\bibitem{pla}
W. Son,  L. Amico, F. Plastina, and V. Vedral, Phys. Rev. A {\bf 79},
022302  (2009).



\bibitem{heinzpeter}
H. P. Breuer and F. Petruccione, {\it The theory of open quantum
systems}  (Oxford University Press, 2002).



\bibitem{Amico06}
L. Amico, F. Baroni, D. Patane', V. Tognetti, and P. Verrucchi, Phys. Rev. A {\bf 74}, 022322 (2006);






\bibitem{haviland} E. Chow, P. Delsing, and D. B. Haviland, Phys. Rev. Lett. {\bf 81}, 204 (1997).

\bibitem{CaiEtAl10} J. Cai, G. G. Guerreschi, and H. J. Briegel, Phys. Rev. Lett. {\bf 104}, 220502(2010).


\end{thebibliography}
\end{document}